\documentclass[conference]{IEEEtran}

\usepackage{cite}
\usepackage{amsmath,amssymb,amsfonts}
\usepackage{algorithmic}
\usepackage{graphicx}
\usepackage{textcomp}
\usepackage{microtype}
\usepackage[hidelinks]{hyperref}
\usepackage{booktabs}
\usepackage{csvsimple}
\usepackage[dvipsnames]{xcolor}

\def\BibTeX{{\rm B\kern-.05em{\sc i\kern-.025em b}\kern-.08em
    T\kern-.1667em\lower.7ex\hbox{E}\kern-.125emX}}

\begin{document}

\title{Simulated Contextual Bandits for Personalization Tasks from Recommendation Datasets}

\author{
    \IEEEauthorblockN{Anton Dereventsov}
    \IEEEauthorblockA{\textit{Lirio AI Research}\\
    \textit{Lirio, LLC}\\
    Knoxville, TN, USA\\
    adereventsov@lirio.com}
\and
    \IEEEauthorblockN{Anton Bibin}
    \IEEEauthorblockA{\textit{Skoltech Agro}\\
    \textit{Skoltech}\\
    Moscow, Russia\\
    a.bibin@skoltech.ru}
}

\maketitle

\begin{abstract}
We propose a method for generating simulated contextual bandit environments for personalization tasks from recommendation datasets like MovieLens, Netflix, Last.fm, Million Song, etc.
This allows for personalization environments to be developed based on real-life data to reflect the nuanced nature of real-world user interactions.
The obtained environments can be used to develop methods for solving personalization tasks, algorithm benchmarking, model simulation, and more.
We demonstrate our approach with numerical examples on MovieLens and IMDb datasets.
\end{abstract}


\section{Introduction}
Personalization tasks consist of learning a user's preferences based on available user features, with the overall goal of providing customized, relevant, and well-received content recommendations.
Motivated by application-oriented platforms like Netflix, YouTube, and Amazon, and encouraged by specialized programs at conferences like ICDM, ACM RecSys, UMAP, and others, the research on recommender systems and personalization techniques has seen a lot of machine learning development in the last decade (see e.g. a survey~\cite{portugal2018use} and the references therein).

In their most conventional formulations, recommendation systems use past interactions as their main source of available user information, and these interactions are often scarce, expensive, and time consuming to collect.
In cases when such information is available, classical approaches like matrix factorization can be deployed to combine the concepts of collaborative and content-based filtering, thus efficiently taking into account all information available to the platform, see e.g.~\cite{koren2009matrix, su2009survey, van2000using}.

Unfortunately, in modern applications it is generally not feasible to collect the required amount of user interactions needed for a non-trivial performance of the recommender system.
However, what modern learning systems do often have is access to some form of user features, which can include but is not limited to the user's interests, demographic data, location, social connections, etc.
Hence there is a need for establishing learning configurations that are capable of leveraging existing data right away, instead of spending time and resources on the collection of new interactions.

Settings where an agent can access a user's features with the goal of learning the user's preferred content are conventionally referred to as \textit{personalization tasks}, see e.g.~\cite{gorgoglione2019recommendation, nabizadeh2020learning, goldenberg2021personalization}.
Examples of personalization applications include web customization~\cite{das2007google, dou2008evaluating}, e-commerce~\cite{goy2007personalization, kaptein2015advancing}, media content~\cite{wei2007video, wang2014exploration}, healthcare~\cite{zhu2018robust, hassouni2018personalization, gomez2022whom}, and many more.

Despite being an actively developing area, research on personalization tasks still lacks established benchmarks that address challenges that are particular to a personalization setting.
And while there exists a wide range of current personalization systems, the collected data is often either proprietary or contains sensitive information on the userbase and thus cannot be publicly released to the research community.
This issue has forced many researchers to design their own simulated personalization environments that model specific applications, see e.g.~\cite{rohde2018recogym, ie2019recsim, dereventsov2021unreasonable}.

In comparison, there are a number of well-established datasets for classical recommendation tasks that contain practical data, such as detailed content descriptions and often real user feedback or ratings for that content.
In this paper we propose utilizing recommendation datasets to generate simulated personalization tasks by forming a contextual bandit environment based on the available data.
We describe an approach that can be applied to a wide variety of conventional datasets and numerically demonstrate it in detail on two particular examples: MovieLens dataset and IMDb dataset.

\section{Preliminaries}
We consider a contextual bandit environment~\cite{langford2007epoch} to model user interactions with the platform/app, as such formulation has become standard in machine learning literature, see e.g.~\cite{li2010contextual, tang2015personalized, lan2016contextual, lei2017actor}.
A contextual bandit environment is represented by the tuple $(\mathcal{S,A},r)$, where $\mathcal{S}$ denotes the state space, $\mathcal{A}$ denotes an action space, and $r$ denotes the reward signal of the environment.

The state space $\mathcal{S}$ indicates the users' parameterizations that the agent observes.
In the case of personalization tasks, a state $s \in \mathcal{S}$ consists of the available user information, such as their demographics, preferences, past interactions, etc.

The action space $\mathcal{A}$ represents the available actions that an agent can take for each state.
Typically each action $a \in \mathcal{A}$ is a parameterization of a piece of content that can be proposed by the agent, i.e. a movie, an article, an advertisement, etc.
Then action parameterization encodes features that are relevant to this type of content, such as tags, genres, runtime, actors/artists, cost, etc.

The reward function $r : \mathcal{S \times A} \to \mathbb{R}$ determines the reward values that the agent receives.
Specifically, $r(s,a) \in \mathbb{R}$ denotes the reward value the agent receives for taking an action $a \in \mathcal{A}$ for a state $s \in \mathcal{S}$.
In practice the reward values are derived from feedback events, which depend on the task at hand but typically include either a user response (liked/disliked), rating ($1$--$5$ stars), engagement (clicked/not clicked), or other similar outcome.
The exact numerical values of the reward function are designed by the practitioners and depend on the possible range of feedback events, type of learning configuration, and other application considerations.
While there are no particular restrictions on the reward design, higher reward values always indicate more preferable outcomes, which the agent is learning to achieve.

An agent deployed on a bandit environment is represented by its policy $\pi : \mathcal{S} \to \mathcal{P(A)}$, where $\mathcal{P(A)}$ denotes the set of probability distributions over the action space $\mathcal{A}$.
A single agent-environment interaction consists of the agent observing a state $s \in \mathcal{S}$, selecting an action $a \in \mathcal{A}$, and receiving a reward signal $r(s,a) \in \mathbb{R}$, i.e.
\[
	s \in \mathcal{S}
	\quad\longrightarrow\quad
	a \in \mathcal{A} \ \big|\ a \sim \pi(s)
	\quad\longrightarrow\quad
	r = r(s,a) \in \mathbb{R}.
\]
Generally after several interactions the agent's policy $\pi$ is updated based on the received reward values.
The frequency of the updates and the exact policy update rule $\pi \leftarrow \pi^\prime$ is specified by the employed reinforcement learning algorithm.
For a more in-depth overview of the field of reinforcement learning we direct the reader to the book~\cite{sutton2018reinforcement} and the references therein.

\section{Proposed Approach}\label{sec:approach}
In this section we develop a method for generating a contextual bandit environment $(\mathcal{S,A},r)$ from a recommendation dataset $(\mathcal{X,Y})$, which usually consists of the user set $\mathcal{X}$ containing the recorded feedback to some of the content, and the content set $\mathcal{Y}$ containing the parameterization of each content item.
On a high level, we will utilize the user set $\mathcal{X}$ to create the state space $\mathcal{S}$ and the content set $\mathcal{Y}$ to create the action space $\mathcal{A}$ by re-parameterizing the available data based on the information provided by the original dataset.
The only component that must be generated from scratch is the reward function $r : \mathcal{S \times A} \to \mathbb{R}$, which can be designed in a variety of ways, depending on the desired outcome.

Before going into more detail on the general approach, we give an illustrative example, considered in~\cite{dudik2011doubly}, where the authors reformulate an image classification problem as a contextual bandit problem.
Consider an image classification dataset $(\mathcal{X,Y})$, e.g. MNIST, CIFAR10, etc., where $\mathcal{X}$ is the set of images and $\mathcal{Y}$ is the set of corresponding labels.
Then it can be written as a contextual bandit environment $(\mathcal{S,A},r)$ in the following way:
\begin{itemize}
	\item the state space $\mathcal{S}$ is the set of input images $\mathcal{X}$;
	\item the action space $\mathcal{A}$ is the set of available labels $\mathcal{Y}$;
	\item the reward function $r : \mathcal{S \times A} \to \mathbb{R}$ is given as
		\[
			r(s,a) = \left\{\begin{array}{ll}
				1 & \text{ if } a \in \mathcal{A} \text{ is a correct label for } s \in \mathcal{S},
				\\
				0 & \text{ otherwise}.
			\end{array}\right.
		\]
\end{itemize}
Despite the differences between classification tasks and contextual bandits, such a setting is regularly utilized in reinforcement learning literature, see e.g.~\cite{swaminathan2015counterfactual, chen2019surrogate, lazic2021optimization}, which further emphasizes a need for a wider range of contextual bandit environments for benchmarking.

\subsection*{Contextual Bandit Construction}
We now state our proposed approach for constructing a contextual bandit environment from a given dataset.
Consider a recommendation dataset $(\mathcal{X,Y})$ consisting of the user set $\mathcal{X}$ and the content set $\mathcal{Y}$.
In order to generate a state space $\mathcal{S}$ and an action space $\mathcal{A}$, we extend the parameterization of the users $x \in \mathcal{X}$ based on their feedback to the available content items $y \in \mathcal{Y}$ and compute the reward signal on this new parameterization.

A content item $y \in \mathcal{Y}$ is typically described by a list of features (tags, genres, artists, etc.) that can serve as a parameterization of an action $a \in \mathcal{A}$ for a contextual bandit formulation after a straightforward re-coding (e.g. one-hot).
Specifically, let $\mathcal{T} = \{t_1, \ldots, t_{|\mathcal{T}|}\}$ be the set of available content features and let $\boldsymbol{e} : \mathcal{T} \to \mathbb{R}^{|\mathcal{T}|}$ indicate the one-hot encoder, i.e. for any $t_j \in \mathcal{T}$
\[
	\boldsymbol{e}(t_j) = (0, \ldots, 0, 1, 0, \ldots, 0) \in \mathbb{R}^{|\mathcal{T}|},
\]
where the value $1$ is on the $j$-th coordinate.
Then for each content item $y \in \mathcal{Y}$ we obtain a parameterization for an action $a \in \mathcal{A}$ by combining the provided features of $y$ as follows:
\begin{equation}\label{eq:action_parameterization}
	a = \sum_{j=1}^{|\mathcal{T}|} \delta(y,t_j) \, \boldsymbol{e}(t_j) \in \mathbb{R}^{|\mathcal{T}|},
\end{equation}
where
\[
	\delta(y,t_j) = \left\{\begin{array}{ll}
		1 & \text{ if content } y \text{ has feature } t_j,
		\\
		0 & \text{ otherwise}.
	\end{array}\right.
\]
Thus a content set $\mathcal{Y}$ with the given features $\mathcal{T}$ provides a parameterization of the action space $\mathcal{A}$.
We note that in recommendation tasks the cardinality of the set $\mathcal{Y}$ might be unfeasibly large from a contextual bandit perspective; in such cases the resulting action space $\mathcal{A}$ can be truncated to obtain a more practical environment.

A user $x \in \mathcal{X}$ is typically represented by their feedback vector $\{f(x,y) \,|\, y \in \mathcal{Y}\}$, where $f(x,y)$ indicates the feedback of user $x$ to the content item $y \in \mathcal{Y}$.
With the content items' parameterization~\eqref{eq:action_parameterization}, we can characterize a state $s \in \mathcal{S}$ through the feedback of user $x \in \mathcal{X}$ as follows:
\begin{equation}\label{eq:state_parameterization}
	s = \sum_{i=1}^{|\mathcal{Y}|} f(x,y_i) \, a_i \in \mathbb{R}^{|\mathcal{T}|},
\end{equation}
where $f(x,y) \in \mathbb{R}$ is the normalized to $[-1,1]$ feedback of user $x$ to the content item $y$, or $0$ if no recorded feedback is available for the pair $(x,y)$.
Thus a user set $\mathcal{X}$, along with available feedback to the content $\mathcal{Y}$, provides a parameterization of the state space $\mathcal{S}$.
We note that even if no user set $\mathcal{X}$ is provided, one can synthesize it by simulating user feedback to the content $\mathcal{Y}$, as we demonstrate in Section~\ref{sec:numerics_imdb}.

Lastly, we establish the reward signal $r : \mathcal{S \times A} \to \mathbb{R}$, which we calculate using the parameterized representation of states and actions, defined in~\eqref{eq:state_parameterization} and~\eqref{eq:action_parameterization} respectively.
In our numerical experiments, for a given pair of the state $s \in \mathcal{S}$ and the action $a \in \mathcal{A}$, the reward value $r(s,a)$ is calculated as the scaled cosine similarity, i.e.
\begin{equation}\label{eq:reward_parameterization}
	r(s,a) = \alpha \frac{\langle s, a \rangle}{\|s\| \|a\|} \in \mathbb{R},
\end{equation}
where $\alpha > 0$ is adjusted to match the feedback distribution $\{f(x,y) \,|\, x \in \mathcal{X},\, y \in \mathcal{Y}\}$ of the original dataset.
In general, there are many other possibilities for designing the reward signal, especially if there are practical considerations needed to match a particular geometry or to satisfy a complexity requirement in order to tailor to a particular application's needs.

Once the parameterizations of the state space $\mathcal{S}$, the action space $\mathcal{A}$, and the reward signal $r : \mathcal{S \times A} \to \mathbb{R}$ are established, the construction of the contextual bandit environment $(\mathcal{S,A},r)$ is complete.
The obtained environment can be used to develop, test, and benchmark various reinforcement learning agents, which we demonstrate in the next section.

We also note that many aspects outlined in the proposed approach, such as re-parameterization of the state and action spaces, or reward signal calculation, can be designed in a variety of ways, depending on the particular aspects required from the simulator.
For concreteness, we state the rules that are utilized in our numerical examples in Section~\ref{sec:numerics}.

\section{Simulated Personalization Environments}\label{sec:numerics}
In this section we demonstrate the process of constructing contextual bandit environments for personalization tasks outlined in Section~\ref{sec:approach}.
Specifically, we consider two well-known recommendation datasets~--- MovieLens 25M and IMDb~--- to generate contextual bandit personalization tasks and deploy reinforcement learning algorithms to solve them.

We note that the main focus of this paper is to propose a way of obtaining personalization tasks from the existing recommendation datasets, and not to train the reinforcement learning agents to solve these environments.
As such, the presented agents serve a purely illustrative purpose and their performance is not necessarily representative of their underlying algorithms.

The presented experiments are performed in Python3 with the use of publicly available libraries.
In particular, the reinforcement learning agents employed in our examples are implemented via Stable-Baselines3\footnote{\url{https://github.com/DLR-RM/stable-baselines3}} library~\cite{raffin2021stable}, and include the following algorithms:
\begin{itemize}
    \item \texttt{A2C}: Advantage Actor Critic~\cite{mnih2016asynchronous};
    \item \texttt{DQN}: Deep Q Network~\cite{mnih2013playing};
    \item \texttt{PPO}: Proximal Policy Optimization~\cite{schulman2017proximal}.
\end{itemize}
The selection of the above algorithms is justified by their wide use in both theoretical research and practical applications.

For technical simplicity, for all agents we use the same $2$-layer policy architecture with $32$ nodes on each layer.
Training is performed via Adam optimizer~\cite{kingma2014adam} with the learning rate of $10^{-4}$.
Otherwise, we use the default values of hyperparameters for each algorithm, as specified by the authors and repository maintainers.

The datasets used in our experiments, as well as the source code are available at~\url{https://github.com/Ansebi/WAIN22_sim_env}.

\subsection{MovieLens 25M Dataset}\label{sec:numerics_ml25m}
MovieLens 25M dataset\footnote{\url{https://grouplens.org/datasets/movielens/25m/}} is a commonly used choice for recommendation tasks, see e.g.~\cite{kuzelewska2014clustering, he2017neural, rendle2020neural, forouzandeh2021presentation}.
The dataset consists of $25,000,000$ ratings to $62,000$ movies by $162,000$ users.
Each movie is tagged with applicable genres, which include [Action, Adventure, Animation, Children's, Comedy, Crime, Documentary, Drama, Fantasy, Film-Noir, Horror, Musical, Mystery, Romance, Sci-Fi, Thriller, War, Western].
Each recorded user has rated at least $20$ movies on a $5$-star scale.
The detailed information on MovieLens 25M dataset is available at~\url{https://files.grouplens.org/datasets/movielens/ml-25m-README.html}.

\begin{figure}[t!]
	\centering
	\includegraphics[width=.49\linewidth]{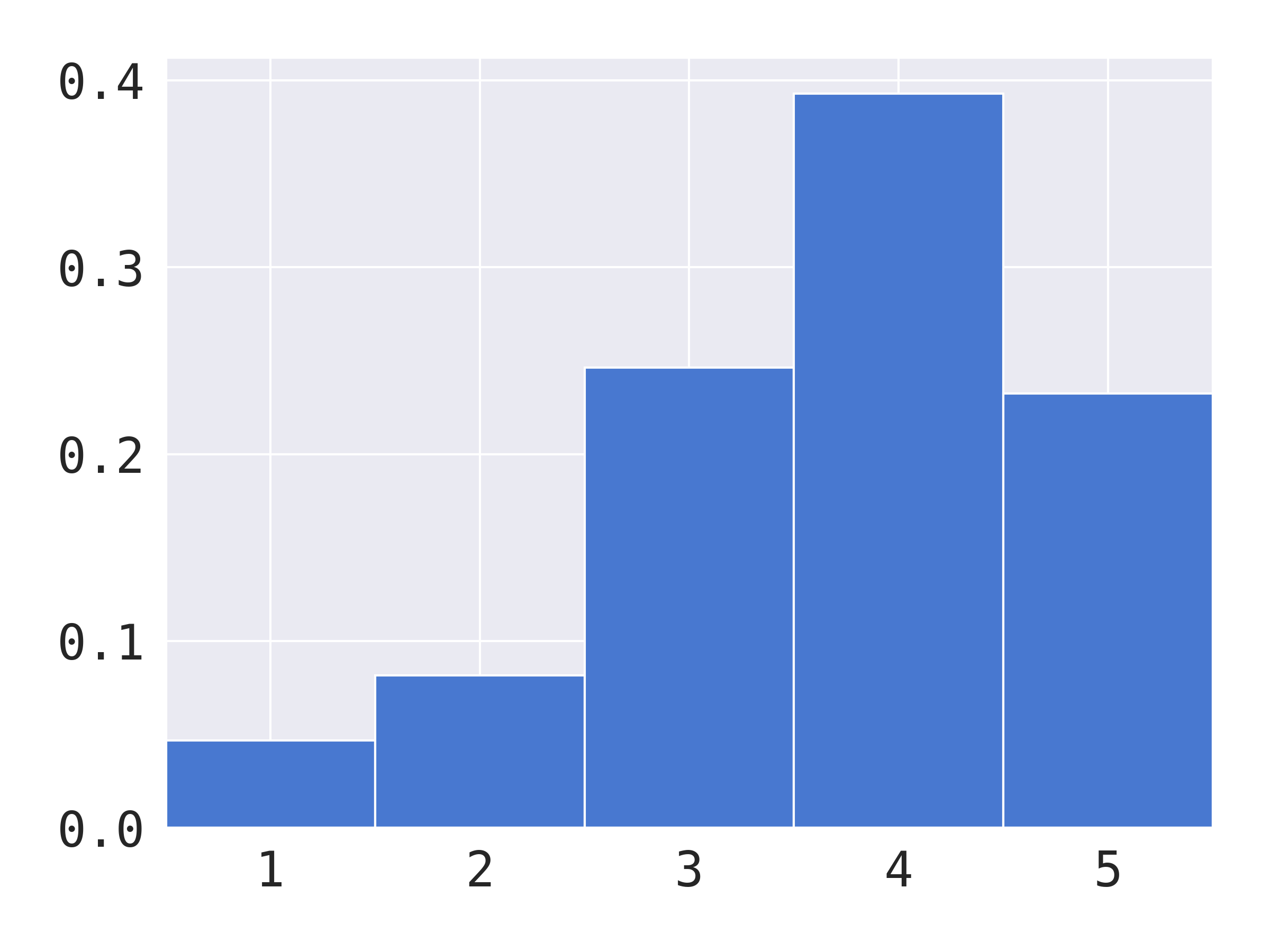}
	\includegraphics[width=.49\linewidth]{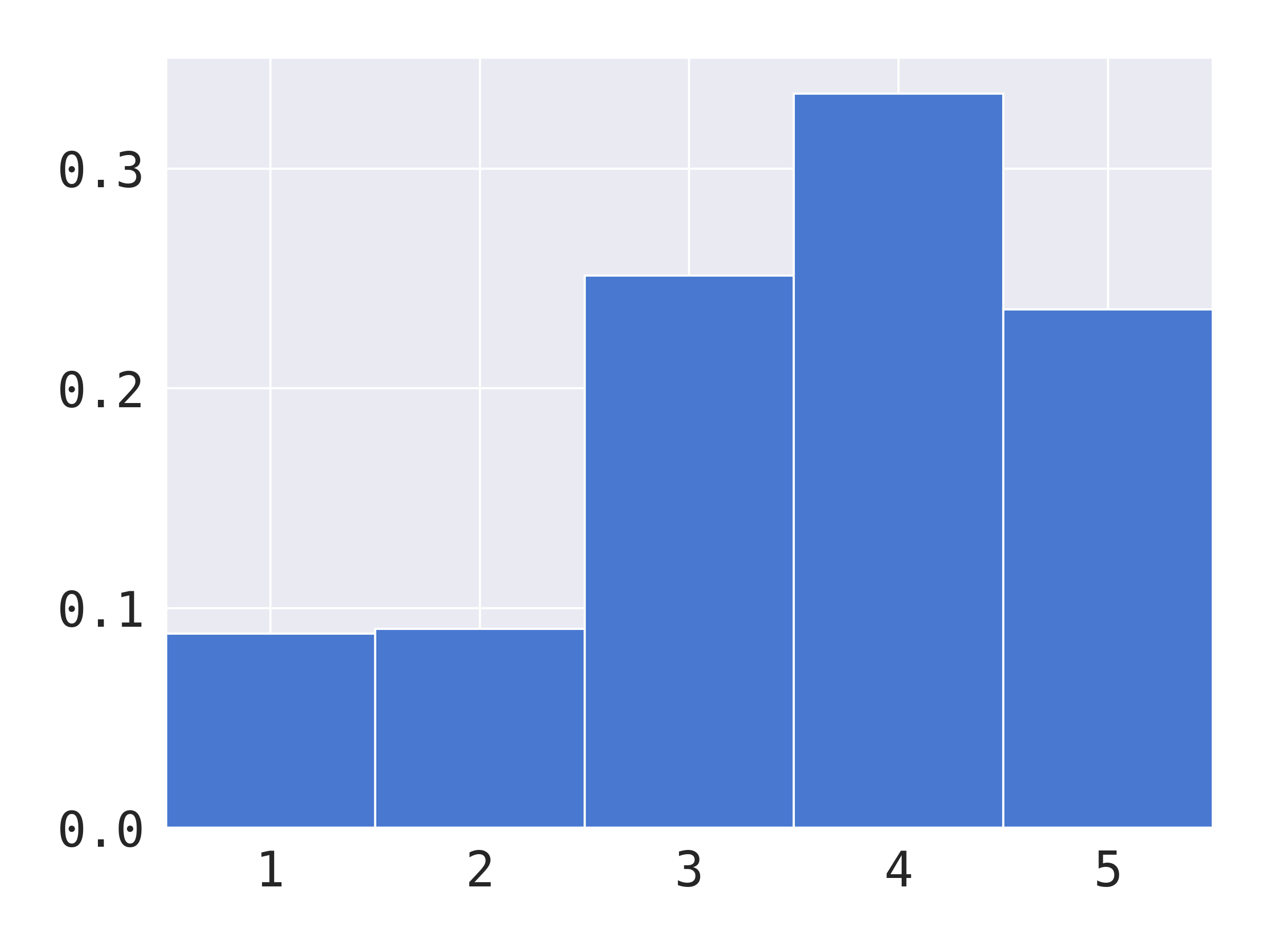}
	\caption{Movie rating distributions: MovieLens 25M dataset (left) and generated contextual bandit environment (right).}
	\label{fig:ml25m_dist}
\end{figure}

\begin{figure}[t!]
	\includegraphics[width=\linewidth]{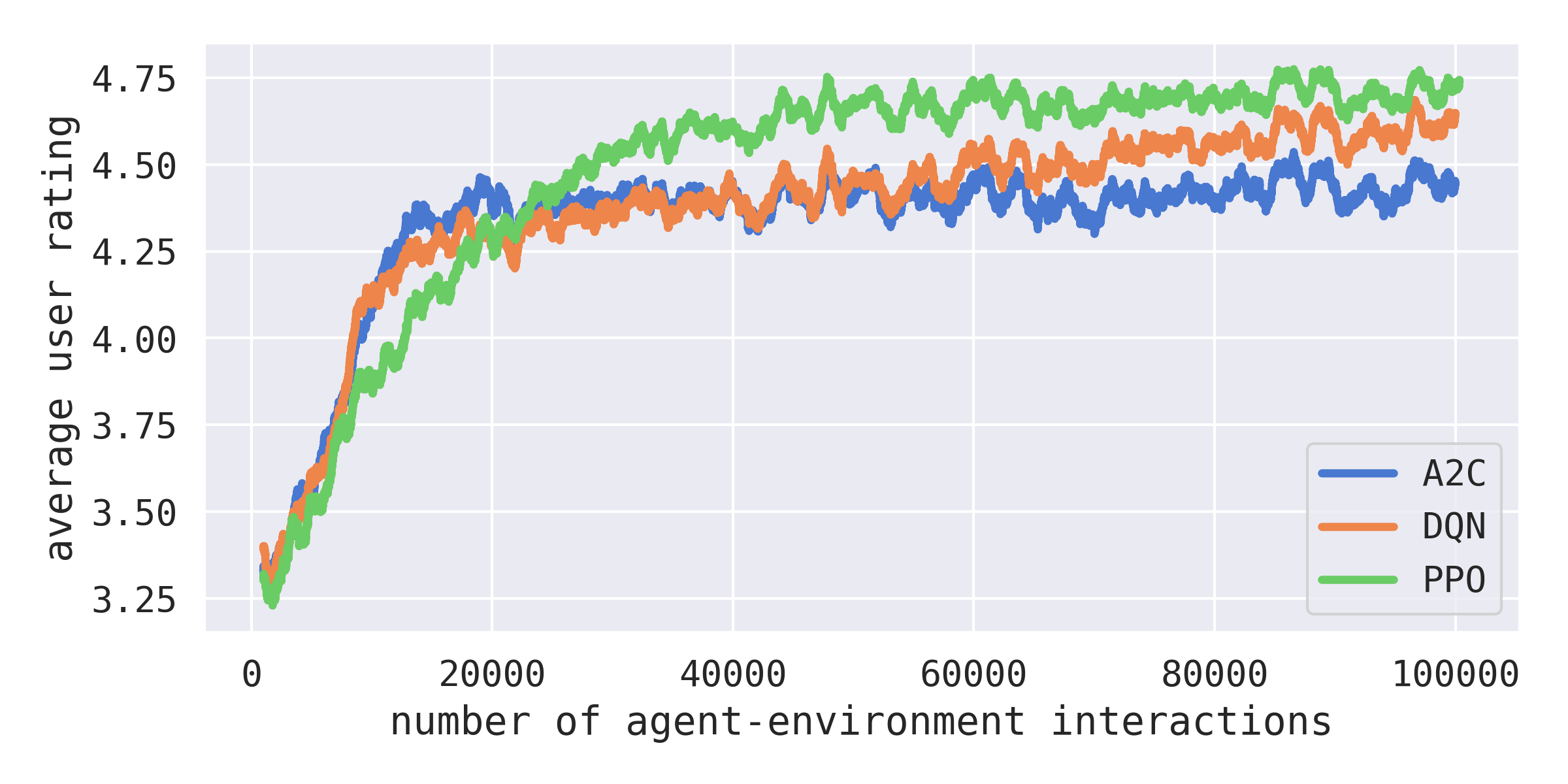}
	\caption{Training rewards achieved by RL agents on the contextual bandit environment generated from MovieLens 25M dataset.}
	\label{fig:ml25m_rewards}
\end{figure}

In this example we use MovieLens 25M dataset to generate the contextual bandit environment $(\mathcal{S,A},r)$ via the approach proposed in Section~\ref{sec:approach}.
In this case the action space $\mathcal{A} \subset \mathbb{R}^{18}$ consists of the $100$ most popular movies, whose parameterizations are obtained via~\eqref{eq:action_parameterization}, see Table~\ref{tab:ml25m_actions}.
The state space $\mathcal{S} \subset \mathbb{R}^{18}$ consists of the $10,000$ most active users, whose features are generated via~\eqref{eq:state_parameterization} with the feedback normalization set to $[0,1]$, see Table~\ref{tab:ml25m_states}.
The reward function $r : \mathcal{S \times A} \to \mathbb{R}$ is constructed via~\eqref{eq:reward_parameterization} with a slight modification to ensure that the reward values align with the original dataset.
Namely, for a user $s \in \mathcal{S}$ and a movie $a \in \mathcal{A}$ the reward is calculated as
\[
	r(s,a) = \operatorname{clip}\left\{
		\frac{1}{2} \left\lceil 2 + 10 \frac{\langle s, a \rangle}{\|s\| \|a\|} \right\rceil,\ 0.5,\ 5.0
	\right\},
\]
where $\lceil x \rceil$ indicates rounding $x$ up to the nearest integer.
This results in the range of possible reward values of $\{0.5, 1.0, 1.5, 2.0, 2.5, 3.0, 3.5, 4.0, 4.5, 5.0\}$, which is the same as the original dataset.
The distribution of reward values is similar to that of the original dataset as presented in Figure~\ref{fig:ml25m_dist}.
The agents' performance on this personalization task is shown in Figure~\ref{fig:ml25m_rewards}.

\begin{table*}
	\scriptsize
	\centering
	\caption{Most popular movies from MovieLens 25M dataset.}
	\begin{filecontents*}{./images/ml25m_actions.csv}
movieId,title,genres,encoding
1,Toy Story (1995),Adventure|Animation|Children|Comedy|Fantasy,[0. 1. 1. 1. 1. 0. 0. 0. 1. 0. 0. 0. 0. 0. 0. 0. 0. 0.]
32,Twelve Monkeys (1995),Mystery|Sci-Fi|Thriller,[0. 0. 0. 0. 0. 0. 0. 0. 0. 0. 0. 0. 1. 0. 1. 1. 0. 0.]
34,Babe (1995),Children|Drama,[0. 0. 0. 1. 0. 0. 0. 1. 0. 0. 0. 0. 0. 0. 0. 0. 0. 0.]
47,Seven (a.k.a. Se7en) (1995),Mystery|Thriller,[0. 0. 0. 0. 0. 0. 0. 0. 0. 0. 0. 0. 1. 0. 0. 1. 0. 0.]
50,Usual Suspects (1995),Crime|Mystery|Thriller,[0. 0. 0. 0. 0. 1. 0. 0. 0. 0. 0. 0. 1. 0. 0. 1. 0. 0.]
110,Braveheart (1995),Action|Drama|War,[1. 0. 0. 0. 0. 0. 0. 1. 0. 0. 0. 0. 0. 0. 0. 0. 1. 0.]
111,Taxi Driver (1976),Crime|Drama|Thriller,[0. 0. 0. 0. 0. 1. 0. 1. 0. 0. 0. 0. 0. 0. 0. 1. 0. 0.]
150,Apollo 13 (1995),Adventure|Drama,[0. 1. 0. 0. 0. 0. 0. 1. 0. 0. 0. 0. 0. 0. 0. 0. 0. 0.]
153,Batman Forever (1995),Action|Adventure|Comedy|Crime,[1. 1. 0. 0. 1. 1. 0. 0. 0. 0. 0. 0. 0. 0. 0. 0. 0. 0.]
165,Die Hard: With a Vengeance (1995),Action|Crime|Thriller,[1. 0. 0. 0. 0. 1. 0. 0. 0. 0. 0. 0. 0. 0. 0. 1. 0. 0.]
\end{filecontents*}
\csvautobooktabular{./images/ml25m_actions.csv}
	\label{tab:ml25m_actions}
\end{table*}

\begin{table*}
	\scriptsize
	\ttfamily
	\centering
	\caption{Most active users from MovieLens 25M dataset.}
	\resizebox{\textwidth}{!}{\begin{filecontents*}{./images/ml25m_states.csv}
userId,encoding
  3,[132.22 ~82.67 ~27.33 20.44 ~~55.11 ~66.67 ~0.56 117.56 ~34.22 ~3.44 ~15.89 ~2.22 ~30.44 ~20.89 ~94.22 100.11 10.89 ~3.56]
 12,[~21.78 ~25.00 ~12.56 ~5.11 ~~74.44 ~30.44 ~5.78 117.22 ~12.67 ~1.78 ~~5.11 10.33 ~10.56 ~48.67 ~~9.44 ~24.44 10.11 ~0.78]
 72,[~30.00 ~20.89 ~~6.78 ~5.67 ~~81.89 ~37.56 ~1.67 120.00 ~17.33 ~5.78 ~17.11 18.44 ~17.56 ~60.78 ~19.00 ~43.22 15.44 ~3.11]
 80,[-11.78 -17.67 ~-1.89 -7.56 ~-48.33 ~12.44 ~0.33 ~-9.22 -13.44 ~0.00 -17.00 -1.56 ~~6.22 -21.00 -20.78 ~~9.11 -2.22 -0.22]
120,[~85.00 ~71.67 ~17.11 25.67 ~169.33 ~67.00 15.89 207.11 ~27.89 ~3.22 ~16.11 17.44 ~27.44 ~82.89 ~45.22 ~83.22 24.00 ~3.56]
166,[~47.78 ~36.78 ~~4.67 11.00 ~~83.56 ~33.11 -0.33 137.33 ~18.44 ~7.44 ~-3.67 20.22 ~22.89 ~97.00 ~15.78 ~51.33 17.00 ~7.00]
171,[106.22 ~80.33 ~18.56 22.22 ~167.67 100.00 ~9.89 224.33 ~38.22 ~4.00 ~23.78 12.33 ~40.56 ~70.44 ~49.67 119.44 25.33 ~7.33]
175,[~24.33 ~~8.44 ~~4.78 ~6.33 ~~47.56 ~27.56 ~1.00 ~90.44 ~~6.00 ~0.33 ~~5.78 ~3.89 ~~8.33 ~37.00 ~~9.56 ~37.22 ~8.33 ~2.78]
181,[-99.89 -64.22 -12.56 -8.11 -111.78 -49.00 -3.22 -82.67 -30.00 -0.78 -46.44 -5.67 -19.78 -35.89 -56.67 -82.67 -6.44 -3.44]
187,[214.00 146.22 ~47.22 46.44 ~247.56 109.44 22.22 266.22 ~85.11 ~1.78 109.44 12.89 ~66.89 ~63.44 145.67 263.67 19.78 ~4.56]
\end{filecontents*}
\csvautobooktabular{./images/ml25m_states.csv}}
	\label{tab:ml25m_states}
\end{table*}

\subsection{IMDb Dataset}\label{sec:numerics_imdb}
Another common source of recommendation datasets is the IMDb repository\footnote{\url{https://datasets.imdbws.com/}} which contains a variety of data commonly used for a range of tasks, see e.g.~\cite{jung2012attribute, oghina2012predicting, yenter2017deep, kumar2019sentiment}.

In this example we use IMDb ``title.basics'' and ``title.ratings'' datasets that contain information on over $9,000,000$ movies, including movie titles, genres, average ratings (on a $10$-star scale), and number of votes.
Detailed information on these datasets is available at~\url{https://www.imdb.com/interfaces/}.
We employ these datasets to construct the contextual bandit environment $(\mathcal{S,A},r)$ via the approach proposed in Section~\ref{sec:approach}.
We note that in this scenario the information on individual user ratings is not available, which we resolve by synthetically simulating user feedback.

We take $10,000$ of the most reviewed movies and their corresponding genres, which in our case consist of ['Action', 'Adventure', 'Animation', 'Biography', 'Comedy', 'Crime', 'Documentary', 'Drama', 'Family', 'Fantasy', 'Film-Noir', 'Game-Show', 'History', 'Horror', 'Music', 'Musical', 'Mystery', 'News', 'Reality-TV', 'Romance', 'Sci-Fi', 'Short', 'Sport', 'Talk-Show', 'Thriller', 'War', 'Western'].
Then each movie is encoded by its genres via~\eqref{eq:action_parameterization} and a user $s \in \mathbb{R}^{27}$ is generated via~\eqref{eq:state_parameterization} with a randomly sampled feedback vector $f \in \mathbb{R}^{10,000}$ containing $50$ non-zero ``user ratings''.
Thus we obtain a parameterization of continuous state space $\mathcal{S} \subset \mathbb{R}^{27}$, representing user preferences, see Table~\ref{tab:imdb_states}.
For the action space $\mathcal{A} \subset \mathbb{R}^{27}$ we take the $100$ most popular movies, see Table~\ref{tab:imdb_actions}.

\begin{figure}[t!]
	\centering
	\includegraphics[width=.49\linewidth]{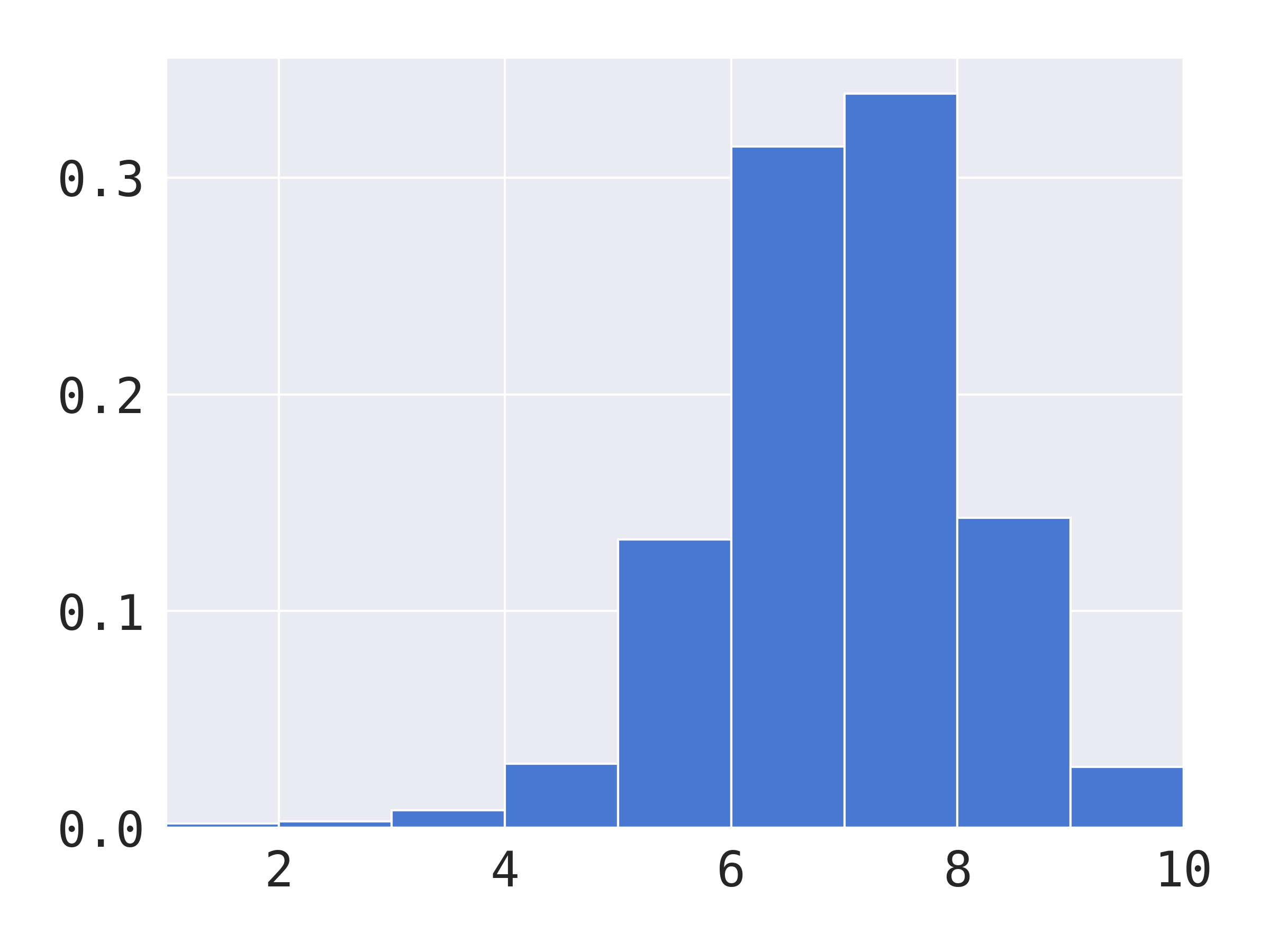}
	\includegraphics[width=.49\linewidth]{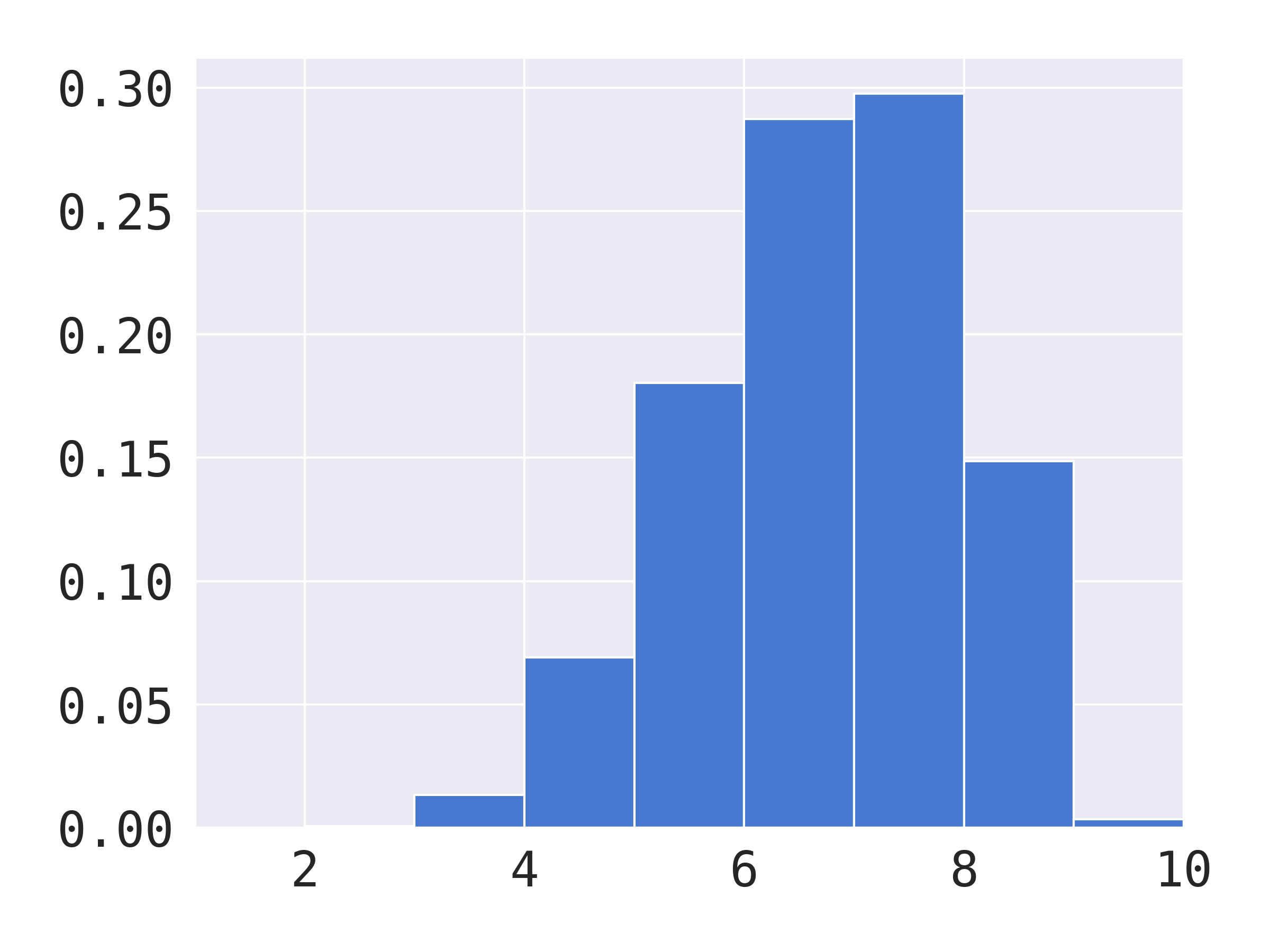}
	\caption{Movie rating distributions: IMDb dataset (left) and generated contextual bandit environment (right).}
	\label{fig:imdb_dist}
\end{figure}

\begin{figure}[t!]
	\includegraphics[width=\linewidth]{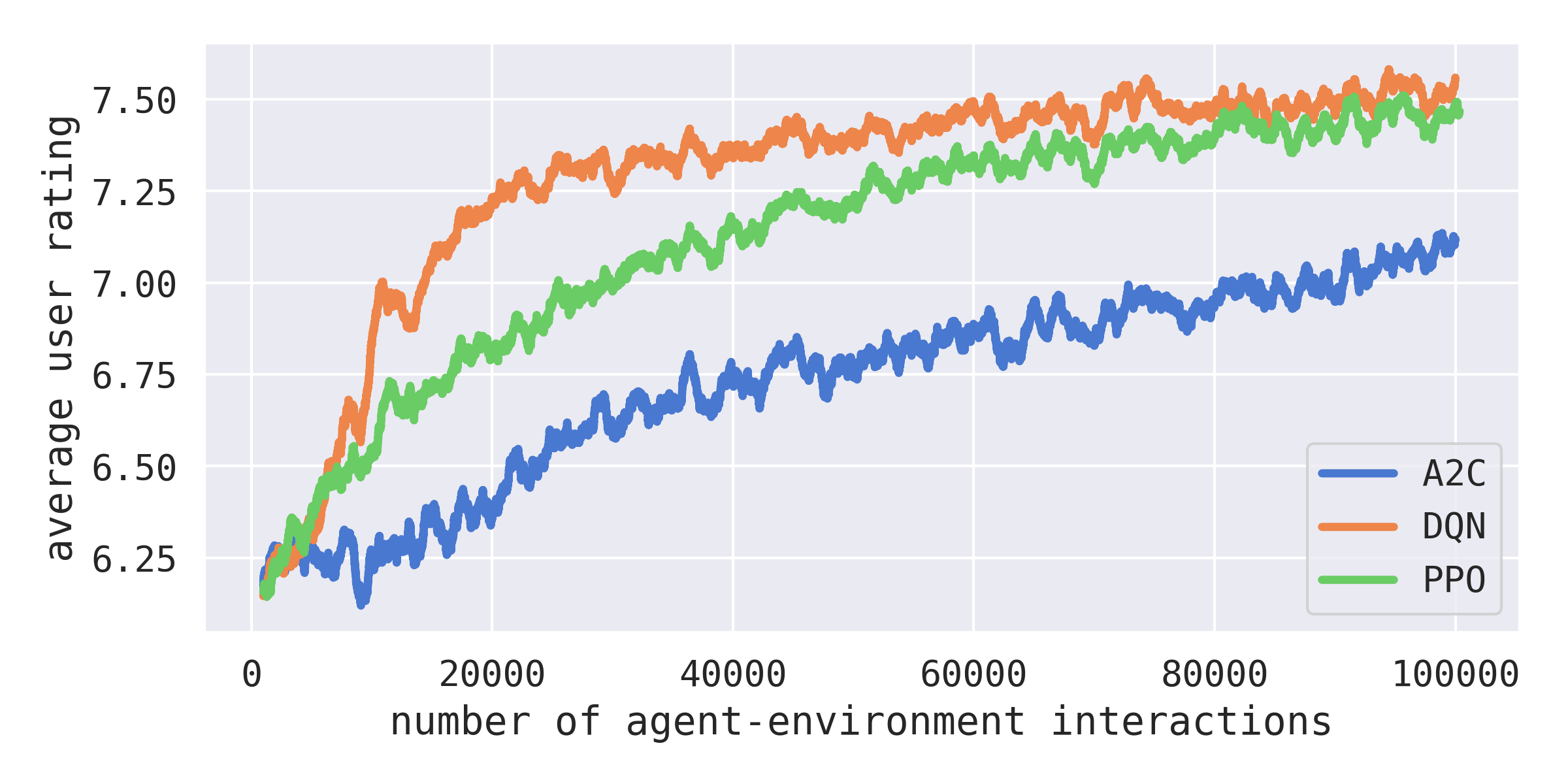}
	\caption{Training rewards achieved by RL agents on the contextual bandit environment generated from IMDb dataset.}
	\label{fig:imdb_rewards}
\end{figure}

The reward function $r : \mathcal{S \times A} \to \mathbb{R}$ is constructed via~\eqref{eq:reward_parameterization} with a slight modification to ensure that the reward values align with the original dataset.
Specifically, for a user $s \in \mathcal{S}$ and a movie $a \in \mathcal{A}$ the reward is calculated as
\[
	r(s,a) = \left\lceil 10 \, \left( \frac{1}{2} + \frac{\langle s, a \rangle}{2 \|s\| \|a\|}\right)^{0.5} \right\rceil,
\]
where $\lceil x \rceil$ indicates rounding $x$ up to the nearest integer.
Then the possible reward values are $\{1, 2, 3, 4, 5, 6, 7, 8, 9, 10\}$, which is the same as in the original dataset.
The distribution of reward values matches that of the original dataset as presented in Figure~\ref{fig:imdb_dist}.
The agents' performance on this personalization task is shown in Figure~\ref{fig:imdb_rewards}.

\begin{table*}
	\scriptsize
	\centering
	\caption{Most popular movies from IMDb dataset.}
	\begin{filecontents*}{./images/imdb_actions.csv}
tconst,primaryTitle,genres,numVotes,encoding
tt0111161,The Shawshank Redemption,Drama,2639333,[0. 0. 0. 0. 0. 0. 0. 1. 0. 0. 0. 0. 0. 0. 0. 0. 0. 0. 0. 0. 0. 0. 0. 0. 0. 0. 0.]
tt0468569,The Dark Knight,Action|Crime|Drama,2610926,[1. 0. 0. 0. 0. 1. 0. 1. 0. 0. 0. 0. 0. 0. 0. 0. 0. 0. 0. 0. 0. 0. 0. 0. 0. 0. 0.]
tt1375666,Inception,Action|Adventure|Sci-Fi,2314403,[1. 1. 0. 0. 0. 0. 0. 0. 0. 0. 0. 0. 0. 0. 0. 0. 0. 0. 0. 0. 1. 0. 0. 0. 0. 0. 0.]
tt0137523,Fight Club,Drama,2083152,[0. 0. 0. 0. 0. 0. 0. 1. 0. 0. 0. 0. 0. 0. 0. 0. 0. 0. 0. 0. 0. 0. 0. 0. 0. 0. 0.]
tt0944947,Game of Thrones,Action|Adventure|Drama,2048783,[1. 1. 0. 0. 0. 0. 0. 1. 0. 0. 0. 0. 0. 0. 0. 0. 0. 0. 0. 0. 0. 0. 0. 0. 0. 0. 0.]
tt0109830,Forrest Gump,Drama|Romance,2043215,[0. 0. 0. 0. 0. 0. 0. 1. 0. 0. 0. 0. 0. 0. 0. 0. 0. 0. 0. 1. 0. 0. 0. 0. 0. 0. 0.]
tt0110912,Pulp Fiction,Crime|Drama,2019908,[0. 0. 0. 0. 0. 1. 0. 1. 0. 0. 0. 0. 0. 0. 0. 0. 0. 0. 0. 0. 0. 0. 0. 0. 0. 0. 0.]
tt0133093,The Matrix,Action|Sci-Fi,1888471,[1. 0. 0. 0. 0. 0. 0. 0. 0. 0. 0. 0. 0. 0. 0. 0. 0. 0. 0. 0. 1. 0. 0. 0. 0. 0. 0.]
tt0068646,The Godfather,Crime|Drama,1829481,[0. 0. 0. 0. 0. 1. 0. 1. 0. 0. 0. 0. 0. 0. 0. 0. 0. 0. 0. 0. 0. 0. 0. 0. 0. 0. 0.]
tt0903747,Breaking Bad,Crime|Drama|Thriller,1824229,[0. 0. 0. 0. 0. 1. 0. 1. 0. 0. 0. 0. 0. 0. 0. 0. 0. 0. 0. 0. 0. 0. 0. 0. 1. 0. 0.]

\end{filecontents*}
\csvautobooktabular{./images/imdb_actions.csv}
	\label{tab:imdb_actions}
\end{table*}

\begin{table*}
	\scriptsize
	\ttfamily
	\centering
	\caption{Users simulated for IMDb dataset.}
	\resizebox{\textwidth}{!}{\begin{filecontents*}{./images/imdb_states.csv}
user,encoding
0,[~2.78 ~2.11 -1.44 -3.22 -5.67 -0.67 ~0.00 -1.44 -2.78 ~1.78 ~0.00 0.00 ~0.00 ~0.78 -1.00 -0.67 ~1.78 0.00 0.00 -0.89 ~0.78 0.00 -1.00 0.00 -0.78 ~0.33 ~0.00]
1,[~3.11 ~3.33 ~1.33 ~0.00 ~1.56 ~1.67 ~0.22 -1.22 ~1.22 ~0.44 ~0.00 0.00 -0.11 ~0.00 ~0.00 -0.11 -1.22 0.00 0.00 -0.67 -2.78 0.33 ~0.00 0.00 ~3.11 ~1.00 ~0.78]
2,[~0.56 ~2.67 -1.67 -1.78 -1.22 ~1.00 ~0.00 ~0.11 -0.67 -1.67 ~0.00 0.00 -0.67 ~0.22 ~1.00 ~0.00 ~2.00 0.00 0.00 -1.78 -0.00 0.00 ~0.00 0.00 ~1.78 -0.22 ~0.00]
3,[-0.44 -1.11 ~0.11 ~0.56 ~0.00 -5.56 -1.00 -2.33 -0.33 ~0.78 ~0.78 0.00 ~0.56 -2.33 -0.67 ~0.00 -2.22 0.00 0.00 ~1.22 ~0.22 0.00 ~0.89 0.00 ~0.11 ~0.00 ~0.44]
4,[~1.89 ~1.67 ~0.11 -0.78 ~1.22 -1.78 ~0.00 -2.56 -1.11 -0.22 ~0.00 0.00 ~0.56 -1.33 ~1.00 ~0.00 -0.78 0.00 0.00 ~3.67 -0.89 0.00 -0.33 0.00 -2.00 ~0.56 -0.78]
5,[~0.56 ~1.44 -0.11 ~0.44 ~0.56 ~1.22 ~0.00 -1.78 -0.22 ~1.22 ~0.00 0.00 ~1.78 ~0.22 -0.33 ~0.00 ~0.44 0.00 0.00 -2.11 -1.56 0.00 ~0.00 0.00 -0.78 ~0.56 ~0.00]
6,[~0.11 -0.33 ~0.56 -2.44 ~3.89 ~0.33 -2.11 -1.22 ~2.78 -1.00 ~0.00 0.00 -2.11 ~0.33 -0.22 ~0.00 ~0.56 0.00 0.11 ~1.89 ~0.22 0.00 -0.78 0.00 ~1.56 ~0.00 ~1.00]
7,[~3.67 ~1.78 -0.67 ~1.00 -1.89 -0.78 -0.11 -2.56 ~1.00 ~1.22 ~0.00 0.00 -0.11 ~1.11 -0.33 ~0.00 ~0.11 0.00 0.00 -1.44 ~1.44 0.00 ~0.00 0.00 -2.56 ~0.56 -0.78]
8,[~1.56 ~2.67 -0.89 ~0.67 -0.33 ~1.44 ~0.11 ~3.22 -0.44 ~0.56 ~0.00 0.00 -0.33 -0.11 -0.78 -0.56 ~2.11 0.00 0.00 ~0.44 ~1.44 0.00 ~0.33 0.00 ~0.89 ~0.33 ~0.00]
9,[-2.44 -0.78 -1.00 -0.78 ~1.22 -0.22 ~0.78 -3.00 -0.22 ~0.22 -1.00 0.00 ~0.00 -1.67 -2.00 -1.00 -4.33 0.00 0.00 ~0.89 ~0.56 0.00 -0.78 0.00 -0.44 -1.00 ~0.00]
\end{filecontents*}
\csvautobooktabular{./images/imdb_states.csv}}
	\label{tab:imdb_states}
\end{table*}

\section{Discussion}
In this paper we propose a method for generating contextual bandit environments from the available recommendation datasets.
Our approach allows for the creation of a variety of configurable personalization tasks to model a wide range of practical challenges and to test various aspects of learning systems.

While the simplicity of the described procedure is beneficial from the implementation and experimentation viewpoints, it also poses fundamental restrictions from an application side.
The main limitation of our approach is that some data-specific information might get lost during the re-parameterization of the state and action spaces, as well as due to the reward function design.
Moreover, the resulting contextual bandit environment is constrained by the original dataset and thus any undesirable patterns contained in the data is likely to be reflected in the simulated environment, which is especially concerning for practical applications.

However, even though our approach is rather simplistic, we hope that it can serve as a step towards the development of community-created personalization benchmarks that assist in the study and advancement of personalization-focused algorithms.
We also note that the proposed methodology can be extended to create more sophisticated simulators that include the domain-specific knowledge required by some practical applications.
In particular, implementing behavioral science insights to model human decision-making into the personalization simulation is an immediate topic of our future research.

\bibliographystyle{abbrv}
\bibliography{references}

\end{document}